\documentclass[prb,preprint,superscriptaddress,pdf]{revtex4-1}
\usepackage{epsfig}
\usepackage{graphicx}
\usepackage{dcolumn}
\usepackage{bm}
\usepackage{tabularx}
\usepackage{hyperref}
\usepackage{xcolor}
\hypersetup{citecolor=black, filecolor= black, linkcolor= black, urlcolor= black}

\begin{document}

\preprint{version 0}

\title{Prediction of stable hafnium carbides: their stoichiometries, mechanical properties, and electronic structure}

\author{Qingfeng Zeng}
\email{qfzeng@nwpu.edu.cn}
\affiliation{Science and Technology on Thermostructural Composite Materials Laboratory, School of Materials Science and Engineering, Northwestern Polytechnical University, Xi'an, Shaanxi 710072, PR China}
\author{Junhui Peng}
\affiliation{Science and Technology on Thermostructural Composite Materials Laboratory, School of Materials Science and Engineering, Northwestern Polytechnical University, Xi'an, Shaanxi 710072, PR China}
\author{Artem R. Oganov}
\affiliation{Department of Geosciences, Center for Materials by Design, and Institute for Advanced Computational Science, State University of New York, Stony Brook, NY 11794-2100, USA}
\affiliation{Moscow Institute of Physics and Technology, Dolgoprudny, Moscow Region 141700, Russia}
\affiliation{School of Materials Science and Engineering, Northwestern Polytechnical University, Xi'an, Shaanxi 710072, PR China}
\author{Qiang Zhu}
\affiliation{Department of Geosciences, Center for Materials by Design, and Institute for Advanced Computational Science, State University of New York, Stony Brook, NY 11794-2100, USA}
\author{Congwei Xie}
\affiliation{Science and Technology on Thermostructural Composite Materials Laboratory, School of Materials Science and Engineering, Northwestern Polytechnical University, Xi'an, Shaanxi 710072, PR China}
\author{Xiaodong Zhang}
\affiliation{Institute of Modern Physics, Northwest University, Xi'an, Shaanxi 710069, PR China}
\affiliation{Science and Technology on Thermostructural Composite Materials Laboratory, School of Materials Science and Engineering, Northwestern Polytechnical University, Xi'an, Shaanxi 710072, PR China}
\author{Dong Dong}
\affiliation{Science and Technology on Thermostructural Composite Materials Laboratory, School of Materials Science and Engineering, Northwestern Polytechnical University, Xi'an, Shaanxi 710072, PR China}
\author{Litong Zhang}
\affiliation{Science and Technology on Thermostructural Composite Materials Laboratory, School of Materials Science and Engineering, Northwestern Polytechnical University, Xi'an, Shaanxi 710072, PR China}
\author{Laifei Cheng}
\affiliation{Science and Technology on Thermostructural Composite Materials Laboratory, School of Materials Science and Engineering, Northwestern Polytechnical University, Xi'an, Shaanxi 710072, PR China}

\begin{abstract}
Hafnium carbides are studied by a systematic search for possible stable stoichiometric compounds in the Hf-C system at ambient pressure using variable-composition \emph{ab initio} evolutionary algorithm implemented in the USPEX code. In addition to well-known HfC, we predicted two additional compounds Hf$_3$C$_2$ and Hf$_6$C$_5$. The structure of Hf$_6$C$_5$ with space group $C2/m$ contains 11 atoms in the primitive cell and this prediction revives the earlier proposal by A. I. Gusev. The stable structure of Hf$_3$C$_2$ also has space group $C2/m$, and is more energetically favorable than the $Immm$, $P\bar3m1$, $P2$ and $C222_1$ structures put forward by A. I. Gusev. Dynamical and mechanical stability of the newly predicted structures have been verified by calculations of their phonons and elastic constants. The bulk and shear moduli of Hf$_3$C$_2$ are 195.8 GPa and 143.1 GPa, respectively, while for Hf$_6$C$_5$ they are 227.9 GPa and 187.2 GPa, respectively. Their mechanical properties are inferior to those of HfC due to the presence of structural vacancies. Chemical bonding, band structure, and Bader charge are presented and discussed.
\end{abstract}
\maketitle

\section{Introduction}

Hafnium carbides, known as ultra-high temperature ceramics, have attracted growing attention because of their unique features. These include extremely high melting temperature and hardness, high thermal and electrical conductivity, and chemical stability, and make them promising advanced materials even in extreme thermal and chemical environments. \cite{LevineSR-JECS-2002,OpekaMM-JMS-2004,SavinoR-AST-2005}

Hafnium carbide is known to crystallize in the NaCl-type structure (space group $Fm\bar3m$) and to have the composition HfC. This is a relatively well-studied material. Its elastic properties and phonon spectra have been studied experimentally\cite{BrownHL-JCP-1966,SmithHG-PRL-1970}, and its structural, elastic, electronic and phonon properties have been computed using first-principles methods\cite{LiH-SSC-2011a,LiH-SSC-2011b}. However, according to the results of theoretical calculation by A. I. Gusev \cite{GusevAI-PSSA-1993}, ordered stoichiometric phases Hf$_3$C$_2$ and Hf$_6$C$_5$ should exist, with possible space groups $Immm$, $P\bar3m1$, $P2$ or $C222_1$ for Hf$_3$C$_2$, and $C2/m$, $P3_1$ or $C2$ for Hf$_6$C$_5$. Experimental synthesis and structure determination of these subtle ordered states encounters problems due to the lack of direct methods. Gusev and Zyryanova \cite{GusevAI-PSSA-2000} studied the order-disorder transition of Hf-C system by measuring the magnetic susceptiblity and confirmed the existence of Hf$_3$C$_2$ and the possible existence of Hf$_6$C$_5$.

Here we explore the stable compounds in the Hf-C system and their crystal structures at ambient pressure using the variable-composition \emph{ab initio} evolutionary algorithm \cite{OganovAR-RMG-2010,OganovAR-ACR-2011}, and discuss their structures, elastic properties and chemical bonding. In Section 2, we describe the computational methods that were used in this work. In Section 3, we present the results - crystal structures, elastic properties and analysis of the electronic structure. Section 4 presents conclusions of this study.

\section{Computational methodology}
The prediction of stable compounds and their crystal structures was performed using evolutionary algorithm implemented in USPEX\cite{OganovAR-JCP-2006,LyaknovAO-CPC-2013} code developed by Oganov's group. This approach features global optimization with real-space representation and flexible physically motivated variation operators. For every candidate structure generated by USPEX, we use first principles structural relaxation, based on density functional theory (DFT) within the Perdew-Burke-Ernzerhof (PBE) generalized gradient approximation (GGA)\cite{PerdewJP-PRL-1996}, as implemented in the VASP code \cite{KresseG-PRB-1996}. The all-electron projector-augmented wave (PAW) method \cite{BlochlPE-PRB-1994}, with the plane-wave kinetic energy cutoff of 900 eV and k-point meshes with reciprocal-space resolution of 2$\pi \times$ 0.07 \AA$^{-1}$ were used. These settings enable excellent convergences of the energy differences, stress tensors and structural parameters.

The calculations of phonon dispersion, elastic properties, band structure, density of state were done using CASTEP code\cite{SegallMD-JPCM-2002} and utilizing the modification of the Perdew-Burke-Ernzerhof GGA functional for solids (PBEsol) \cite{PerdewJP-PRL-2008}. Norm-conserving scheme \cite{HamannDR-PRL-1979} is used to generate pseudopotentials for Hf and C with the choices of electronic configuration of [Xe]6s$^2$5d$^2$, and [He]2s$^2$2p$^2$, respectively. In these plane-wave calculations, the cutoff energy is 750 eV, and k-point separation is 2$\pi \times$ 0.05 \AA$^{-1}$. Bader charge analysis\cite{BaderR-1990} was performed using Bader Charge Analysis code\cite{TandW-JPCM-2009,SanvilleSD-JCC-2007,HenkelmanG-CMS-2006}.

\section{Results and discussions}
\subsection{Crystal structure prediction and structural properties}
Variable-composition evolutionary algorithm\cite{OganovAR-RMG-2010,OganovAR-ACR-2011} used in this work is very effective in simultaneously predicting stable compositions and their structures for multi-component systems. In our searches, we allowed all possible compositions in the Hf-C system with structures containing up to 30 atoms in the unit cell. The initial generation consisted of 50 structures, and all subsequent generations had 40 structures.  50\% of new structures were produced by heredity, 20\% by softmutation, 10\% by transmutation, 20\% by random symmetric structure generator. Stable structures and their compositions were determined using the convex hull construction: a compound is thermodynamically stable if the enthalpy of its decomposition into any other compounds is positive.

\begin{figure}
\includegraphics[width=\linewidth]{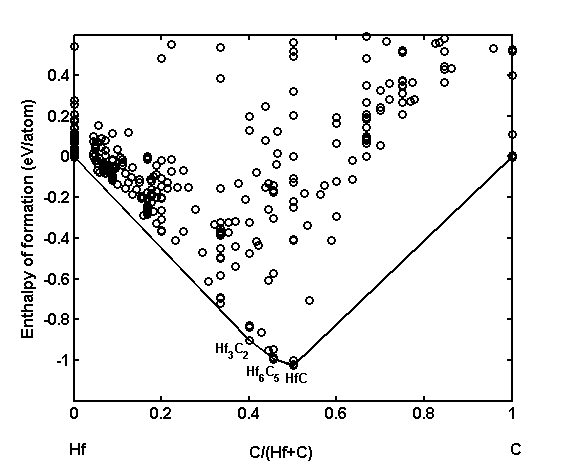}
\caption{\label{Fig1}Convex hull of Hf-C system at ambient pressure. The circle presents different structure and those structures locating on lines are thermodynamically stable.}
\end{figure}

We explored the possible stable crystal structures for Hf-C system at ambient pressure and zero kelvin. In addition to rocksalt-type (B1) HfC (space group $Fm\bar3m$), we found another two compounds Hf$_3$C$_2$ and Hf$_6$C$_5$, both belonging to space group $C2/m$. The enthalpies of formation of the predicted structures are shown in Fig. \ref{Fig1}. It can be clearly seen that HfC, Hf$_6$C$_5$ and Hf$_3$C$_2$ are thermodynamically stable compounds.

\begin{figure}
\includegraphics[width=0.5\linewidth]{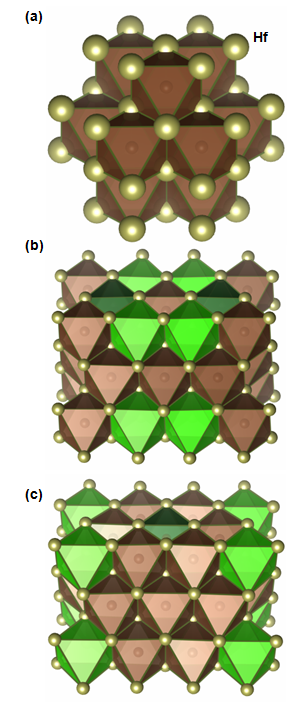}
\caption{\label{Fig2}The crystal structures of (a) HfC with one unit cell, and (b) Hf$_3$C$_2$ (c) Hf$_6$C$_5$ with twice unit cells. The space group of HfC, Hf$_3$C$_2$ and Hf$_6$C$_5$ is $Fm\bar3m$,$C2/m$, $C2/m$, respectively.}
\end{figure}

The crystal structures of Hf$_3$C$_2$ and Hf$_6$C$_5$ are shown in Fig. \ref{Fig2}, and their crystallographic data and enthalpies of formation are listed in Table \ref{Tab1}. For comparison, we also present computational results on HfC in Table \ref{Tab1}. Structure of Hf$_3$C$_2$ has space group of $C2/m$ and 20 atoms in the conventional unit cell (Fig. \ref{Fig2}b). Two more structures of Hf$_3$C$_2$ ($Immm$ and $P\bar3m1$) proposed by A. I. Gusev\cite{GusevAI-PSSA-1993} were also found during our searches. However, their enthalpies are higher than that of $C2/m$, which is therefore more stable. Crystal structure of Hf$_6$C$_5$ is shown in Fig. \ref{Fig2}c. Its space group is also $C2/m$ and its structure has 22 atoms in the conventional unit cell, its structural parameters are presented in Table \ref{Tab1}, and agree with the theoretical calculation of A. I. Gusev\cite{GusevAI-PSSA-1993}.

\begin{table}
\centering
\caption{Crystallographic data and enthalpies of formation of Hf$_3$C$_2$ and Hf$_6$C$_5$, the data of HfC is also presented as a comparison}
\begin{tabular}{ c c  c c c  c }   
\hline \hline
Compound   & Space group  & Volume  &  Lattice constants& Atom position & Enthalpy of formation \\
 & (No.) &(\AA$^3$/f.u.)  &(\AA) & (Wyckoff position) &  (eV/atom) \\
\hline
Hf$_3$C$_2$ & $C2/m$ & 75.91 & $a$=5.720 & Hf(4i) (0.737, 0.500, 0.744) & -0.901 \\
 & (12)  &  & $b$=9.893 &Hf(8j) (0.248, 0.161, 0.264) & \\
 &          &  & $c$=5.701 &C(2a) (0.0, 0.0, 0.0) & \\
 &          &  &                   &C(2d) (0.0, 0.5, 0.5) & \\
 &          &  &                   &C(4h) (0.0, 0.834, 0.5) & \\

Hf$_6$C$_5$ & $C2/m$ & 153.14 & $a$=5.729 & Hf(4i) (0.739, 0.000, 0.260) & -0.995 \\
 & (12)  &  & $b$=9.900 &Hf(8j) (0.241, 0.327, 0.746) & \\
 &          &  & $c$=5.731 &C(4g) (0.0, 0.333, 0.0) & \\
 &          &  &                   &C(2d) (0.0, 0.5, 0.5) & \\
 &          &  &                   &C(4h) (0.0, 0.832, 0.5) & \\

 HfC & $Fm\bar{3}m$ & 25.34 & $a$=4.675 & Hf(4a) (0.0, 0.00, 0.00) & -1.027 \\
 & (225)  &  & $a$=4.637\cite{LiH-SSC-2011b} &C(4b) (0.5, 0.5, 0.5)  & \\
 &          &  & $a$=4.639\cite{Nartowski-JMC-1999} & & \\

\hline
\end{tabular}
\label{Tab1}
\end{table}

All the stable hafnium carbides are strongly related structures and can be derived from HfC structure with the highest symmetry. HfC has a structure of cubic-packing hafnium atoms, and carbon atoms fill all octahedral voids(Fig. \ref{Fig2}a), which is an ideal cubic rocksalt-type structure. The octahedra shown in Fig. \ref{Fig2}b and c with green color are empty, i.e. formed by six Hf atoms but without interstitial C atoms. In Hf$_3$C$_2$ structure, only 2/3 of carbon octahedral voids are filled (and 1/3 are vacant), and in Hf$_6$C$_5$ 5/6 are filled (and 1/6 are vacant). In both Hf$_3$C$_2$ and Hf$_6$C$_5$, the vacancies appear in every second octahedral layer, with 1/3 of in-layer octahedra occupied (Hf$_3$C$_2$) or 2/3 of in-layer octahedra occupied (Hf$_6$C$_5$). Ordering of the vacancies in both cases lowers the symmetry from cubic ($Fm\bar3m$) to monoclinic ($C2/m$). Moreover, due to the vacancies, the coordination number of Hf atoms varies in different systems: 6 in HfC, 5 in Hf$_6$C$_5$, and 4 in Hf$_3$C$_2$, while in all these structures carbon atoms invariably had the coordination number 6 (octahedral coordination). In this way, Hf$_6$C$_5$ and Hf$_3$C$_2$ can be described as defective rocksalt-type structures.

It is instructive to look at molecular volumes (see Table \ref{Tab1}). The volumes per formula unit (f.u.) of Hf$_3$C$_2$ (75.91 \AA$^3$/f.u.), Hf$_6$C$_5$ (153.14 \AA$^3$/f.u.) and HfC (25.536 \AA$^3$/f.u.) correspond to practically constant volume per Hf atom (25.3-25.5 \AA$^3$ per Hf atom). Fig. \ref{Fig1} shows that the most prominent stable state is HfC. And Hf$_3$C$_2$ and especially Hf$_6$C$_5$ will be stable only in a narrow range of chemical potentials in hafnium-rich conditions. This explains why HfC is well known from experiment, while Hf$_3$C$_2$ and especially Hf$_6$C$_5$ are more elusive.

\begin{figure}
\centering
\begin{minipage}{0.5\textwidth}
\includegraphics[scale=0.8]{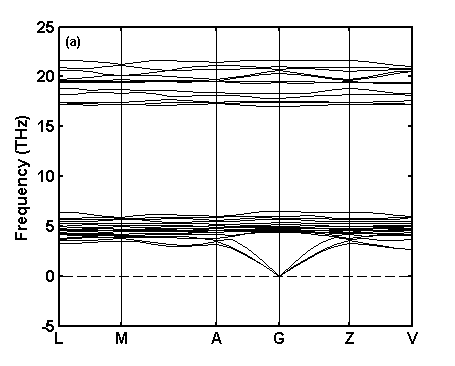}
\end{minipage}
\begin{minipage}{0.5\textwidth}
\includegraphics[scale=0.8]{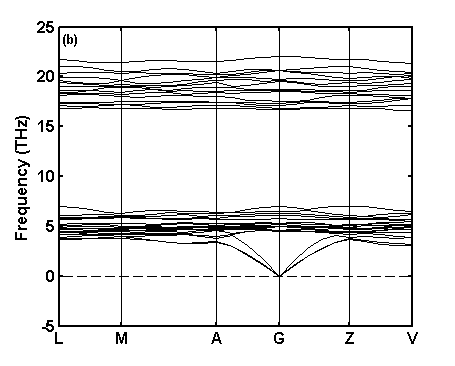}
\end{minipage}
\caption{\label{Fig3}Phonon dispersion curves of (a) Hf$_3$C$_2$ (b) Hf$_6$C$_5$ at ambient pressure}
\end{figure}

To verify the dynamical stability of the newly predicted Hf$_3$C$_2$ and Hf$_6$C$_5$, we computed their phonon dispersions (Fig. \ref{Fig3}). No imaginary phonon frequencies were found throughout the Brillouin zone, suggesting dynamical stability of these phases.

\subsection{Elastic properties}
The elastic constants of a material describe its response to an applied stress or, conversely, the stress required to maintain a given deformation, and can be used to evaluate the elastic properties. These properties are directly related to mechanical stability. The criteria of mechanical stability of a monoclinic crystal are as follows\cite{CowleyRA-PRB-1976}:
\begin{equation}\label{eqK2}
K_2 = \det \left| {C_{ij}} \right|,\;\;i,j \le 5,\;\; K_2 > 0, \;\;C_{44}C_{66} - C_{46}^2  > 0
\end{equation}

The calculated elastic constants of Hf$_3$C$_2$ and Hf$_6$C$_5$ at the ground state are listed in Table \ref{Tab2}. It is obvious that these criteria are satisfied, suggesting that Hf$_3$C$_2$ and Hf$_6$C$_5$ are mechanically stable.

The bulk modulus characterizes the response of a material to volume change, $B = P/(\Delta V/V)$. $P$ is the applied pressure and $\Delta V$ is the volume change. The shear modulus characterizes the response to shear deformation, $G = \tau /\gamma$. $\tau$ is the shear stress and $\gamma$ is shear strain. The bulk modulus $B$  and shear modulus $G$ can be obtained from elastic constants \cite{HillR-PRSA-1952}, and their values for Hf$_3$C$_2$ and Hf$_6$C$_5$ are presented in Table \ref{Tab2}. For comparison, the values of the elastic properties of HfC are also presented in Table \ref{Tab2}, and one can see a good agreement with the values reported in literature. The bulk and shear moduli of Hf$_3$C$_2$ and Hf$_6$C$_5$ are lower than those of HfC. The Pugh's ratios\cite{PughSF-PM-1954} of $G_H/B_H$ of the three compounds are larger than 0.57, indicating brittleness of these compounds. According to Eq. \ref{eqHv}\cite{ChenXQ-IM-2011}, the Vickers hardness of Hf$_3$C$_2$, Hf$_6$C$_5$, and HfC is 22.28, 30.91 and 32.95 GPa, respectively - lowering of the hardness from HfC to Hf$_6$C$_5$ to Hf$_3$C$_2$ is an expected consequence of vacancies.

\begin{equation}\label{eqHv}
H_V = 2 * (k^2*G)^{0.585}-3
\end{equation}

\begin{table}
\caption{Calculated elastic constants $C_{ij}$ , the bulk modulus, shear modulus and hardness (GPa) of Hf$_3$C$_2$, Hf$_6$C$_5$ and HfC at the ground state, and some literature values of HfC}
\centering
\begin{tabular}{lll lll l}
\hline \hline
Compound & Hf$_3$C$_2$ & Hf$_6$C$_5$ &  &  HfC & &\\
         &             &             & This work & Calc.\cite{HeL-SM-2008} & Expt. \cite{WeberW-PRB-1973} & Expt. \cite{BrownHL-JCP-1966}\\
\hline
$C_{11}$ & 391 & 448 & 589 & 577 & 500 & \\
$C_{22}$ & 418 & 471 & & & & \\
$C_{33}$ & 372 & 470 & & & & \\
$C_{44}$ & 126 & 181 & 192 & 171 & 180 & \\
$C_{55}$ & 153 & 213 & & & & \\
$C_{66}$ & 143 & 199 & & & & \\
$C_{12}$ & 198 & 118 & 98 & 117 & & \\
$C_{13}$ & 102 & 117 & & & & \\
$C_{15}$ & -9.7 & -1.4 & & & & \\
$C_{23}$ & 91 & 96 & & & & \\
$C_{25}$ & 12 & 28 & & & & \\
$C_{35}$ & -0.6 & -21 & & & & \\
$C_{46}$ & -7.2 & 26 & & & & \\
$B_H$ & 195.8 & 227.9 & 261.7 & 270 & & 242 \\
$G_H$ & 143.1 & 187.2 & 212.2 & 230 & & 195\\
$k$\footnote {Pugh's ratio: $k=G_H/B_H$} & 0.73 & 0.82 & 0.81 & & & \\
$H_V$ & 22.28 & 30.91 & 32.95 & & & \\
\hline
\end{tabular}
\label{Tab2}
\end{table}

\subsection{Chemical bonding}
\begin{figure}
\begin{minipage}{0.5\textwidth}
\scalebox{1.2}[1.2]{\includegraphics[scale=0.3]{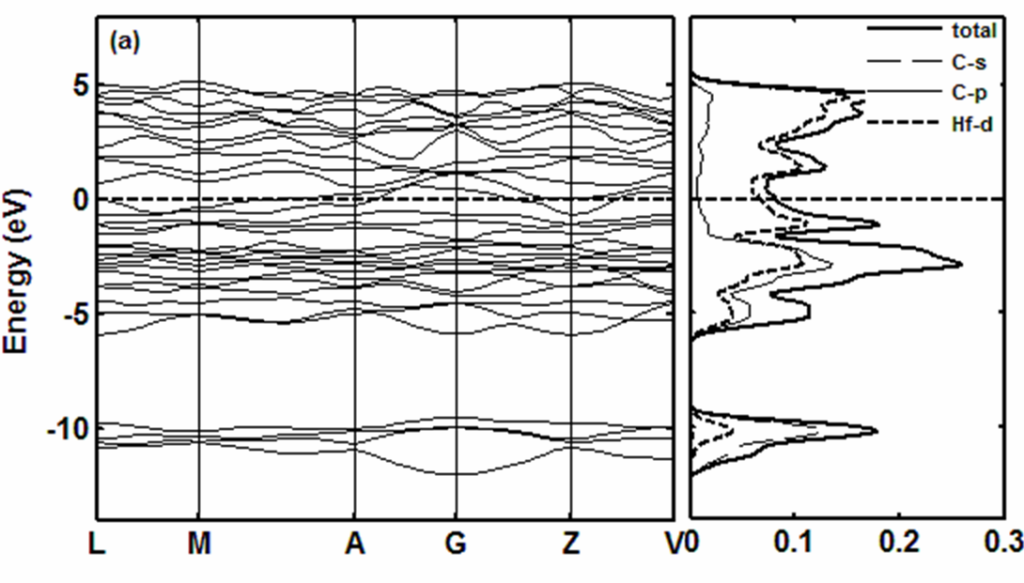}}
\end{minipage}

\begin{minipage}{0.5\textwidth}
\scalebox{1.2}[1.2]{\includegraphics[scale=0.3]{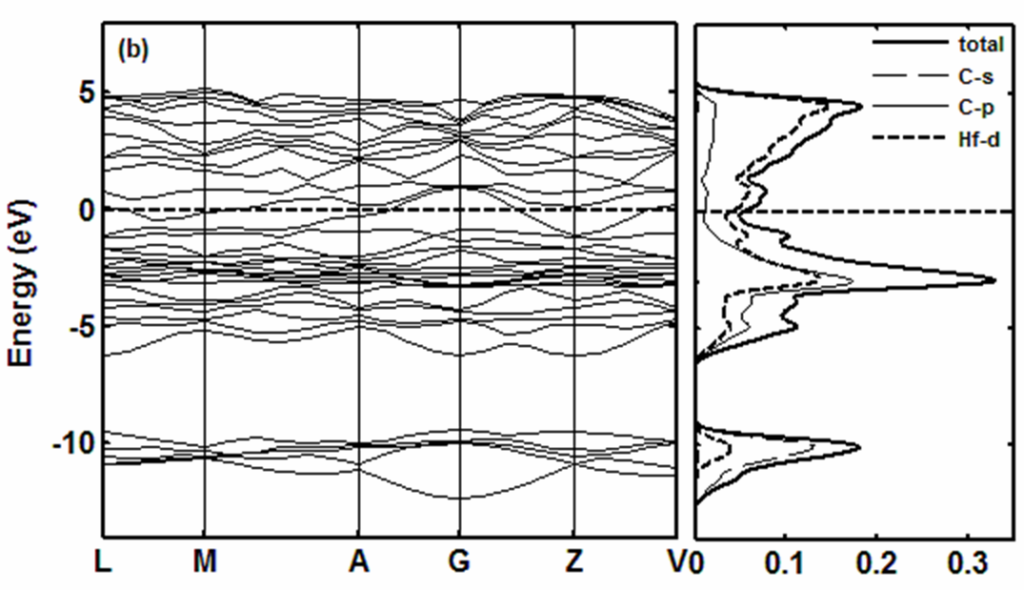}}
\end{minipage}

\begin{minipage}{0.5\textwidth}
\scalebox{1.2}[1.2]{\includegraphics[scale=0.3]{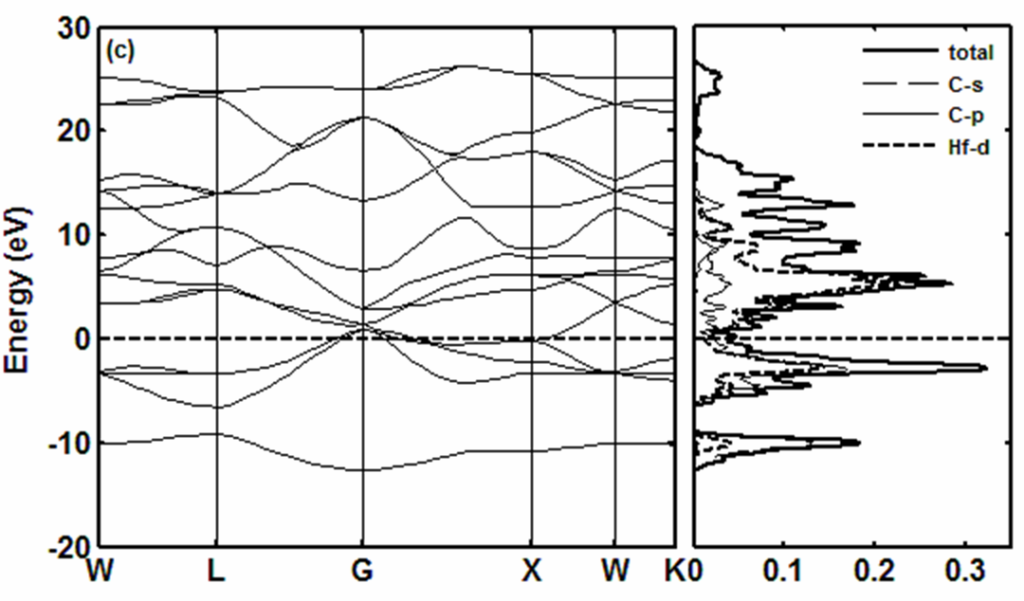}}
\end{minipage}

\caption{\label{Fig4}The band structure and density of states of (a) Hf$_3$C$_2$ (b) Hf$_6$C$_5$ (c) HfC}
\end{figure}

The band structures and DOS of these three compounds are shown in Fig. \ref{Fig4}. All the stable hafnium carbides are weak metals, as seen from finite but small DOS at the Fermi level - 3.28, 2.41, and 0.312 electrons/eV for Hf$_3$C$_2$, Hf$_6$C$_5$ and HfC, respectively. Taking into account the number of atoms in the primitive cells of Hf$_3$C$_2$, Hf$_6$C$_5$ and HfC, which were used in the calculations of the band structure and DOS, we see that the DOS at the Fermi level normalized per valence electron decreases as the number of vacancies decreases - from Hf$_3$C$_2$ (0.082 states/eV) to Hf$_6$C$_5$ (0.054 states/eV) and to HfC (0.039 states/eV). In all these three compounds there are pronounced pseudogaps at the Fermi level, and bonding can be characterized as mixed metallic-covalent. Indeed, orbital-projected DOS indicates strong hybridization of C-$p$ and Hf-$d$ valence states below the Fermi energy, i.e. presence of significant covalency in all three compounds. Bader charge analysis show that each Hf atom gives 1.734 electrons to each C atom in HfC. In case of Hf$_6$C$_5$, Hf atoms contribute 1.521$\pm0.013$ electrons/atom, and C atoms get 1.826$\pm0.019$ electrons/atom. In case of Hf$_3$C$_2$, Hf atoms contribute 1.267$\pm0.026$ electrons/atom, and C atoms get 1.90$\pm0.044$ electrons/atom. This, in agreement with the DOS, shows lower metallicity of HfC compared that of Hf$_6$C$_5$ and Hf$_3$C$_2$. Thus, HfC could have the highest hardness and melting point among these three compounds.

\section{Conclusions}
In this paper we explored the possible stable compounds and structures in the Hf-C system at ambient pressure using variable-composition evolutionary algorithm. Besides the well-known HfC ($Fm\bar3m$), another two stoichiometric compounds (Hf$_3$C$_2$, $C2/m$ and Hf$_6$C$_5$, $C2/m$) are found. All three stable hafnium carbides have rocksalt-type structures: HfC with ideal cubic structure without vacancies, and defective rocksalt-type phases Hf$_3$C$_2$ and Hf$_6$C$_5$ with monoclinic symmetry ($C2/m$) due to ordering of carbon vacancies. Their elastic constants and phonon dispersions are also calculated, which verify their mechanical and dynamical stabilities. Their bulk modulus and shear modulus are lower than that of HfC due to the presence of structural vacancies. We found that all three compounds are weak metals, with increasing metallicity as the concentration of vacancies increases. We also found significant covalency in all these weakly metallic compounds, while their hardness values fall with increasing concentration of vacancies.

\begin{acknowledgments}
We thank the Research Fund of the State Key Laboratory of Solidification Processing of NWPU(No. 65-TP-2011), the Basic Research Foundation of NWPU (No. JCY20130114), the Foreign Talents Introduction and Academic Exchange Program (No. B08040), the National Science Foundation (Nos. EAR-1114313, DMR-1231586), DARPA (Nos. W31P4Q1310005, W31P4Q1210008), and the Government of the Russian Federation (No. 14.A12.31.0003) for financial support. The authors also acknowledge the High Performance Computing Center of NWPU for the allocation of computing time on their machines. USPEX code, with options for global optimization of the thermodynamic potential (energy, enthalpy, free energy), hardness, bandgap, dielectric constant, and other properties, is available at: http://uspex.stonybrook.edu.
\end{acknowledgments}

\bibliographystyle{apsrev4-1}
\bibliography{biblio}

\providecommand{\noopsort}[1]{}\providecommand{\singleletter}[1]{#1}%
\begin{thebibliography}{30}%
\makeatletter
\providecommand \@ifxundefined [1]{%
 \@ifx{#1\undefined}
}%
\providecommand \@ifnum [1]{%
 \ifnum #1\expandafter \@firstoftwo
 \else \expandafter \@secondoftwo
 \fi
}%
\providecommand \@ifx [1]{%
 \ifx #1\expandafter \@firstoftwo
 \else \expandafter \@secondoftwo
 \fi
}%
\providecommand \natexlab [1]{#1}%
\providecommand \enquote  [1]{``#1''}%
\providecommand \bibnamefont  [1]{#1}%
\providecommand \bibfnamefont [1]{#1}%
\providecommand \citenamefont [1]{#1}%
\providecommand \href@noop [0]{\@secondoftwo}%
\providecommand \href [0]{\begingroup \@sanitize@url \@href}%
\providecommand \@href[1]{\@@startlink{#1}\@@href}%
\providecommand \@@href[1]{\endgroup#1\@@endlink}%
\providecommand \@sanitize@url [0]{\catcode `\\12\catcode `\$12\catcode
  `\&12\catcode `\#12\catcode `\^12\catcode `\_12\catcode `\%12\relax}%
\providecommand \@@startlink[1]{}%
\providecommand \@@endlink[0]{}%
\providecommand \url  [0]{\begingroup\@sanitize@url \@url }%
\providecommand \@url [1]{\endgroup\@href {#1}{\urlprefix }}%
\providecommand \urlprefix  [0]{URL }%
\providecommand \Eprint [0]{\href }%
\providecommand \doibase [0]{http://dx.doi.org/}%
\providecommand \selectlanguage [0]{\@gobble}%
\providecommand \bibinfo  [0]{\@secondoftwo}%
\providecommand \bibfield  [0]{\@secondoftwo}%
\providecommand \translation [1]{[#1]}%
\providecommand \BibitemOpen [0]{}%
\providecommand \bibitemStop [0]{}%
\providecommand \bibitemNoStop [0]{.\EOS\space}%
\providecommand \EOS [0]{\spacefactor3000\relax}%
\providecommand \BibitemShut  [1]{\csname bibitem#1\endcsname}%
\let\auto@bib@innerbib\@empty
\bibitem [{\citenamefont {Levine}\ \emph {et~al.}(2002)\citenamefont {Levine},
  \citenamefont {Opila}, \citenamefont {Halbig}, \citenamefont {Kiser},
  \citenamefont {Singh},\ and\ \citenamefont {Salem}}]{LevineSR-JECS-2002}%
  \BibitemOpen
  \bibfield  {author} {\bibinfo {author} {\bibfnamefont {S.~R.}\ \bibnamefont
  {Levine}}, \bibinfo {author} {\bibfnamefont {E.~J.}\ \bibnamefont {Opila}},
  \bibinfo {author} {\bibfnamefont {M.~C.}\ \bibnamefont {Halbig}}, \bibinfo
  {author} {\bibfnamefont {J.~D.}\ \bibnamefont {Kiser}}, \bibinfo {author}
  {\bibfnamefont {M.}~\bibnamefont {Singh}}, \ and\ \bibinfo {author}
  {\bibfnamefont {J.~A.}\ \bibnamefont {Salem}},\ }\href@noop {} {\bibfield
  {journal} {\bibinfo  {journal} {J. Eur. Ceram. Soc.}\ }\textbf {\bibinfo
  {volume} {22}},\ \bibinfo {pages} {2757} (\bibinfo {year}
  {2002})}\BibitemShut {NoStop}%
\bibitem [{\citenamefont {Opeka}\ \emph {et~al.}(2004)\citenamefont {Opeka},
  \citenamefont {Talmy},\ and\ \citenamefont {Zaykosk}}]{OpekaMM-JMS-2004}%
  \BibitemOpen
  \bibfield  {author} {\bibinfo {author} {\bibfnamefont {M.~M.}\ \bibnamefont
  {Opeka}}, \bibinfo {author} {\bibfnamefont {I.~G.}\ \bibnamefont {Talmy}}, \
  and\ \bibinfo {author} {\bibfnamefont {J.~A.}\ \bibnamefont {Zaykosk}},\
  }\href@noop {} {\bibfield  {journal} {\bibinfo  {journal} {J. Mater. Sci.}\
  }\textbf {\bibinfo {volume} {39}},\ \bibinfo {pages} {5887} (\bibinfo {year}
  {2004})}\BibitemShut {NoStop}%
\bibitem [{\citenamefont {Savino}\ \emph {et~al.}(2005)\citenamefont {Savino},
  \citenamefont {Fumo}, \citenamefont {Paterna},\ and\ \citenamefont
  {Serpico}}]{SavinoR-AST-2005}%
  \BibitemOpen
  \bibfield  {author} {\bibinfo {author} {\bibfnamefont {R.}~\bibnamefont
  {Savino}}, \bibinfo {author} {\bibfnamefont {M.~D.~S.}\ \bibnamefont {Fumo}},
  \bibinfo {author} {\bibfnamefont {D.}~\bibnamefont {Paterna}}, \ and\
  \bibinfo {author} {\bibfnamefont {M.}~\bibnamefont {Serpico}},\ }\href@noop
  {} {\bibfield  {journal} {\bibinfo  {journal} {Aerosp. Sci. Technol.}\
  }\textbf {\bibinfo {volume} {9}},\ \bibinfo {pages} {151} (\bibinfo {year}
  {2005})}\BibitemShut {NoStop}%
\bibitem [{\citenamefont {Brown}\ \emph {et~al.}(1966)\citenamefont {Brown},
  \citenamefont {Armstrong},\ and\ \citenamefont {Kempter}}]{BrownHL-JCP-1966}%
  \BibitemOpen
  \bibfield  {author} {\bibinfo {author} {\bibfnamefont {H.~L.}\ \bibnamefont
  {Brown}}, \bibinfo {author} {\bibfnamefont {P.~E.}\ \bibnamefont
  {Armstrong}}, \ and\ \bibinfo {author} {\bibfnamefont {C.~P.}\ \bibnamefont
  {Kempter}},\ }\href@noop {} {\bibfield  {journal} {\bibinfo  {journal} {J.
  Chem. Phys.}\ }\textbf {\bibinfo {volume} {45}},\ \bibinfo {pages} {547}
  (\bibinfo {year} {1966})}\BibitemShut {NoStop}%
\bibitem [{\citenamefont {Smith}\ and\ \citenamefont
  {Gl\"aser}(1970)}]{SmithHG-PRL-1970}%
  \BibitemOpen
  \bibfield  {author} {\bibinfo {author} {\bibfnamefont {H.~G.}\ \bibnamefont
  {Smith}}\ and\ \bibinfo {author} {\bibfnamefont {W.}~\bibnamefont
  {Gl\"aser}},\ }\href {\doibase 10.1103/PhysRevLett.25.1611} {\bibfield
  {journal} {\bibinfo  {journal} {Phys. Rev. Lett.}\ }\textbf {\bibinfo
  {volume} {25}},\ \bibinfo {pages} {1611} (\bibinfo {year}
  {1970})}\BibitemShut {NoStop}%
\bibitem [{\citenamefont {Li}\ \emph {et~al.}(2011{\natexlab{a}})\citenamefont
  {Li}, \citenamefont {Zhang}, \citenamefont {Zeng}, \citenamefont {Ren},
  \citenamefont {Guan}, \citenamefont {Liu},\ and\ \citenamefont
  {Cheng}}]{LiH-SSC-2011a}%
  \BibitemOpen
  \bibfield  {author} {\bibinfo {author} {\bibfnamefont {H.}~\bibnamefont
  {Li}}, \bibinfo {author} {\bibfnamefont {L.}~\bibnamefont {Zhang}}, \bibinfo
  {author} {\bibfnamefont {Q.}~\bibnamefont {Zeng}}, \bibinfo {author}
  {\bibfnamefont {H.}~\bibnamefont {Ren}}, \bibinfo {author} {\bibfnamefont
  {K.}~\bibnamefont {Guan}}, \bibinfo {author} {\bibfnamefont {Q.}~\bibnamefont
  {Liu}}, \ and\ \bibinfo {author} {\bibfnamefont {L.}~\bibnamefont {Cheng}},\
  }\href@noop {} {\bibfield  {journal} {\bibinfo  {journal} {Solid State
  Commun.}\ }\textbf {\bibinfo {volume} {151}},\ \bibinfo {pages} {61}
  (\bibinfo {year} {2011}{\natexlab{a}})}\BibitemShut {NoStop}%
\bibitem [{\citenamefont {Li}\ \emph {et~al.}(2011{\natexlab{b}})\citenamefont
  {Li}, \citenamefont {Zhang}, \citenamefont {Zeng}, \citenamefont {Guan},
  \citenamefont {Li}, \citenamefont {Ren}, \citenamefont {Liu},\ and\
  \citenamefont {Cheng}}]{LiH-SSC-2011b}%
  \BibitemOpen
  \bibfield  {author} {\bibinfo {author} {\bibfnamefont {H.}~\bibnamefont
  {Li}}, \bibinfo {author} {\bibfnamefont {L.}~\bibnamefont {Zhang}}, \bibinfo
  {author} {\bibfnamefont {Q.}~\bibnamefont {Zeng}}, \bibinfo {author}
  {\bibfnamefont {K.}~\bibnamefont {Guan}}, \bibinfo {author} {\bibfnamefont
  {K.}~\bibnamefont {Li}}, \bibinfo {author} {\bibfnamefont {H.}~\bibnamefont
  {Ren}}, \bibinfo {author} {\bibfnamefont {S.}~\bibnamefont {Liu}}, \ and\
  \bibinfo {author} {\bibfnamefont {L.}~\bibnamefont {Cheng}},\ }\href@noop {}
  {\bibfield  {journal} {\bibinfo  {journal} {Solid State Commun.}\ }\textbf
  {\bibinfo {volume} {151}},\ \bibinfo {pages} {602} (\bibinfo {year}
  {2011}{\natexlab{b}})}\BibitemShut {NoStop}%
\bibitem [{\citenamefont {Gusev}\ and\ \citenamefont
  {Rempel}(1993)}]{GusevAI-PSSA-1993}%
  \BibitemOpen
  \bibfield  {author} {\bibinfo {author} {\bibfnamefont {A.~I.}\ \bibnamefont
  {Gusev}}\ and\ \bibinfo {author} {\bibfnamefont {A.~A.}\ \bibnamefont
  {Rempel}},\ }\href@noop {} {\bibfield  {journal} {\bibinfo  {journal} {Phys.
  Stat. Sol. (a)}\ }\textbf {\bibinfo {volume} {135}},\ \bibinfo {pages} {15}
  (\bibinfo {year} {1993})}\BibitemShut {NoStop}%
\bibitem [{\citenamefont {Gusev}\ and\ \citenamefont
  {Zyryanova}(2000)}]{GusevAI-PSSA-2000}%
  \BibitemOpen
  \bibfield  {author} {\bibinfo {author} {\bibfnamefont {A.~I.}\ \bibnamefont
  {Gusev}}\ and\ \bibinfo {author} {\bibfnamefont {A.~N.}\ \bibnamefont
  {Zyryanova}},\ }\href {\doibase
  10.1002/(SICI)1521-396X(200002)177:2<419::AID-PSSA419>3.0.CO;2-J} {\bibfield
  {journal} {\bibinfo  {journal} {Phys. Stat. Sol. (a)}\ }\textbf {\bibinfo
  {volume} {177}},\ \bibinfo {pages} {419} (\bibinfo {year}
  {2000})}\BibitemShut {NoStop}%
\bibitem [{\citenamefont {Oganov}\ \emph {et~al.}(2010)\citenamefont {Oganov},
  \citenamefont {Ma}, \citenamefont {Lyakhov}, \citenamefont {Valle},\ and\
  \citenamefont {Gatti}}]{OganovAR-RMG-2010}%
  \BibitemOpen
  \bibfield  {author} {\bibinfo {author} {\bibfnamefont {A.}~\bibnamefont
  {Oganov}}, \bibinfo {author} {\bibfnamefont {Y.}~\bibnamefont {Ma}}, \bibinfo
  {author} {\bibfnamefont {A.}~\bibnamefont {Lyakhov}}, \bibinfo {author}
  {\bibfnamefont {M.}~\bibnamefont {Valle}}, \ and\ \bibinfo {author}
  {\bibfnamefont {C.}~\bibnamefont {Gatti}},\ }\href@noop {} {\bibfield
  {journal} {\bibinfo  {journal} {Rev. Mineral. Geochem.}\ }\textbf {\bibinfo
  {volume} {71}},\ \bibinfo {pages} {271} (\bibinfo {year} {2010})}\BibitemShut
  {NoStop}%
\bibitem [{\citenamefont {Oganov}\ \emph {et~al.}(2011)\citenamefont {Oganov},
  \citenamefont {Lyakhov},\ and\ \citenamefont {Valle}}]{OganovAR-ACR-2011}%
  \BibitemOpen
  \bibfield  {author} {\bibinfo {author} {\bibfnamefont {A.~R.}\ \bibnamefont
  {Oganov}}, \bibinfo {author} {\bibfnamefont {A.~O.}\ \bibnamefont {Lyakhov}},
  \ and\ \bibinfo {author} {\bibfnamefont {M.}~\bibnamefont {Valle}},\
  }\href@noop {} {\bibfield  {journal} {\bibinfo  {journal} {Accounts Chem.
  Res.}\ }\textbf {\bibinfo {volume} {44}},\ \bibinfo {pages} {227} (\bibinfo
  {year} {2011})}\BibitemShut {NoStop}%
\bibitem [{\citenamefont {Oganov}\ and\ \citenamefont
  {Glass}(2006)}]{OganovAR-JCP-2006}%
  \BibitemOpen
  \bibfield  {author} {\bibinfo {author} {\bibfnamefont {A.~R.}\ \bibnamefont
  {Oganov}}\ and\ \bibinfo {author} {\bibfnamefont {C.~W.}\ \bibnamefont
  {Glass}},\ }\href@noop {} {\bibfield  {journal} {\bibinfo  {journal} {J.
  Chem. Phys.}\ }\textbf {\bibinfo {volume} {124}},\ \bibinfo {pages} {244704}
  (\bibinfo {year} {2006})}\BibitemShut {NoStop}%
\bibitem [{\citenamefont {Lyakhov}\ \emph {et~al.}(2013)\citenamefont
  {Lyakhov}, \citenamefont {Oganov}, \citenamefont {Stokes},\ and\
  \citenamefont {Zhu}}]{LyaknovAO-CPC-2013}%
  \BibitemOpen
  \bibfield  {author} {\bibinfo {author} {\bibfnamefont {A.~O.}\ \bibnamefont
  {Lyakhov}}, \bibinfo {author} {\bibfnamefont {A.~R.}\ \bibnamefont {Oganov}},
  \bibinfo {author} {\bibfnamefont {H.~T.}\ \bibnamefont {Stokes}}, \ and\
  \bibinfo {author} {\bibfnamefont {Q.}~\bibnamefont {Zhu}},\ }\href@noop {}
  {\bibfield  {journal} {\bibinfo  {journal} {Comput. Phys. Commun.}\ }\textbf
  {\bibinfo {volume} {184}},\ \bibinfo {pages} {1172} (\bibinfo {year}
  {2013})}\BibitemShut {NoStop}%
\bibitem [{\citenamefont {Perdew}\ \emph {et~al.}(1996)\citenamefont {Perdew},
  \citenamefont {Burke},\ and\ \citenamefont {Ernzerhof}}]{PerdewJP-PRL-1996}%
  \BibitemOpen
  \bibfield  {author} {\bibinfo {author} {\bibfnamefont {J.~P.}\ \bibnamefont
  {Perdew}}, \bibinfo {author} {\bibfnamefont {K.}~\bibnamefont {Burke}}, \
  and\ \bibinfo {author} {\bibfnamefont {M.}~\bibnamefont {Ernzerhof}},\
  }\href@noop {} {\bibfield  {journal} {\bibinfo  {journal} {Phys. Rev. Lett.}\
  }\textbf {\bibinfo {volume} {77}},\ \bibinfo {pages} {3865} (\bibinfo {year}
  {1996})}\BibitemShut {NoStop}%
\bibitem [{\citenamefont {Kresse}\ and\ \citenamefont
  {Furthm\"{u}ller}(1996)}]{KresseG-PRB-1996}%
  \BibitemOpen
  \bibfield  {author} {\bibinfo {author} {\bibfnamefont {G.}~\bibnamefont
  {Kresse}}\ and\ \bibinfo {author} {\bibfnamefont {J.}~\bibnamefont
  {Furthm\"{u}ller}},\ }\href@noop {} {\bibfield  {journal} {\bibinfo
  {journal} {Phys. Rev. B}\ }\textbf {\bibinfo {volume} {54}},\ \bibinfo
  {pages} {11169} (\bibinfo {year} {1996})}\BibitemShut {NoStop}%
\bibitem [{\citenamefont {Bl\"ochl}(1994)}]{BlochlPE-PRB-1994}%
  \BibitemOpen
  \bibfield  {author} {\bibinfo {author} {\bibfnamefont {P.~E.}\ \bibnamefont
  {Bl\"ochl}},\ }\href@noop {} {\bibfield  {journal} {\bibinfo  {journal}
  {Phys. Rev. B}\ }\textbf {\bibinfo {volume} {50}},\ \bibinfo {pages} {17953}
  (\bibinfo {year} {1994})}\BibitemShut {NoStop}%
\bibitem [{\citenamefont {Segall}\ \emph {et~al.}(2002)\citenamefont {Segall},
  \citenamefont {Lindan}, \citenamefont {Probert}, \citenamefont {Pickard},
  \citenamefont {Hasnip}, \citenamefont {Clark},\ and\ \citenamefont
  {Payne}}]{SegallMD-JPCM-2002}%
  \BibitemOpen
  \bibfield  {author} {\bibinfo {author} {\bibfnamefont {M.~D.}\ \bibnamefont
  {Segall}}, \bibinfo {author} {\bibfnamefont {P.~J.~D.}\ \bibnamefont
  {Lindan}}, \bibinfo {author} {\bibfnamefont {M.~J.}\ \bibnamefont {Probert}},
  \bibinfo {author} {\bibfnamefont {C.~J.}\ \bibnamefont {Pickard}}, \bibinfo
  {author} {\bibfnamefont {P.~J.}\ \bibnamefont {Hasnip}}, \bibinfo {author}
  {\bibfnamefont {S.~J.}\ \bibnamefont {Clark}}, \ and\ \bibinfo {author}
  {\bibfnamefont {M.}~\bibnamefont {Payne}},\ }\href@noop {} {\bibfield
  {journal} {\bibinfo  {journal} {J. Phys.: Condens. Matter}\ }\textbf
  {\bibinfo {volume} {14}},\ \bibinfo {pages} {2717} (\bibinfo {year}
  {2002})}\BibitemShut {NoStop}%
\bibitem [{\citenamefont {Perdew}\ \emph {et~al.}(2008)\citenamefont {Perdew},
  \citenamefont {Ruzsinszky}, \citenamefont {Csonka}, \citenamefont {Vydrov},
  \citenamefont {Scuseria}, \citenamefont {Constantin}, \citenamefont {Zhou},\
  and\ \citenamefont {Burke}}]{PerdewJP-PRL-2008}%
  \BibitemOpen
  \bibfield  {author} {\bibinfo {author} {\bibfnamefont {J.~P.}\ \bibnamefont
  {Perdew}}, \bibinfo {author} {\bibfnamefont {A.}~\bibnamefont {Ruzsinszky}},
  \bibinfo {author} {\bibfnamefont {G.~I.}\ \bibnamefont {Csonka}}, \bibinfo
  {author} {\bibfnamefont {O.~A.}\ \bibnamefont {Vydrov}}, \bibinfo {author}
  {\bibfnamefont {G.~E.}\ \bibnamefont {Scuseria}}, \bibinfo {author}
  {\bibfnamefont {L.~A.}\ \bibnamefont {Constantin}}, \bibinfo {author}
  {\bibfnamefont {X.}~\bibnamefont {Zhou}}, \ and\ \bibinfo {author}
  {\bibfnamefont {K.}~\bibnamefont {Burke}},\ }\href@noop {} {\bibfield
  {journal} {\bibinfo  {journal} {Phys. Rev. Lett.}\ }\textbf {\bibinfo
  {volume} {100}},\ \bibinfo {pages} {6406} (\bibinfo {year}
  {2008})}\BibitemShut {NoStop}%
\bibitem [{\citenamefont {Hamann}\ \emph {et~al.}(1979)\citenamefont {Hamann},
  \citenamefont {Schluter},\ and\ \citenamefont {Chiang}}]{HamannDR-PRL-1979}%
  \BibitemOpen
  \bibfield  {author} {\bibinfo {author} {\bibfnamefont {D.~R.}\ \bibnamefont
  {Hamann}}, \bibinfo {author} {\bibfnamefont {M.}~\bibnamefont {Schluter}}, \
  and\ \bibinfo {author} {\bibfnamefont {C.}~\bibnamefont {Chiang}},\
  }\href@noop {} {\bibfield  {journal} {\bibinfo  {journal} {Phys. Rev. Lett.}\
  }\textbf {\bibinfo {volume} {43}},\ \bibinfo {pages} {1494} (\bibinfo {year}
  {1979})}\BibitemShut {NoStop}%
\bibitem [{\citenamefont {Bader}(1990)}]{BaderR-1990}%
  \BibitemOpen
  \bibfield  {author} {\bibinfo {author} {\bibfnamefont {R.}~\bibnamefont
  {Bader}},\ }\href@noop {} {\emph {\bibinfo {title} {Atoms in Molecules: A
  Quantum Theory}}}\ (\bibinfo  {publisher} {University Press},\ \bibinfo
  {address} {New York},\ \bibinfo {year} {1990})\BibitemShut {NoStop}%
\bibitem [{\citenamefont {Tang}\ \emph {et~al.}(2009)\citenamefont {Tang},
  \citenamefont {Sanville},\ and\ \citenamefont {Henkelman}}]{TandW-JPCM-2009}%
  \BibitemOpen
  \bibfield  {author} {\bibinfo {author} {\bibfnamefont {W.}~\bibnamefont
  {Tang}}, \bibinfo {author} {\bibfnamefont {E.}~\bibnamefont {Sanville}}, \
  and\ \bibinfo {author} {\bibfnamefont {G.}~\bibnamefont {Henkelman}},\
  }\href@noop {} {\bibfield  {journal} {\bibinfo  {journal} {J. Phys.: Condens.
  Matter}\ }\textbf {\bibinfo {volume} {21}},\ \bibinfo {pages} {084204}
  (\bibinfo {year} {2009})}\BibitemShut {NoStop}%
\bibitem [{\citenamefont {Sanville}\ \emph {et~al.}(2007)\citenamefont
  {Sanville}, \citenamefont {Kenny}, \citenamefont {Smith},\ and\ \citenamefont
  {Henkelman}}]{SanvilleSD-JCC-2007}%
  \BibitemOpen
  \bibfield  {author} {\bibinfo {author} {\bibfnamefont {E.}~\bibnamefont
  {Sanville}}, \bibinfo {author} {\bibfnamefont {S.~D.}\ \bibnamefont {Kenny}},
  \bibinfo {author} {\bibfnamefont {R.}~\bibnamefont {Smith}}, \ and\ \bibinfo
  {author} {\bibfnamefont {G.}~\bibnamefont {Henkelman}},\ }\href@noop {}
  {\bibfield  {journal} {\bibinfo  {journal} {J. Comp. Chem.}\ }\textbf
  {\bibinfo {volume} {28}},\ \bibinfo {pages} {899} (\bibinfo {year}
  {2007})}\BibitemShut {NoStop}%
\bibitem [{\citenamefont {Henkelman}\ \emph {et~al.}(2006)\citenamefont
  {Henkelman}, \citenamefont {Arnaldsson},\ and\ \citenamefont
  {J\'{o}nsson}}]{HenkelmanG-CMS-2006}%
  \BibitemOpen
  \bibfield  {author} {\bibinfo {author} {\bibfnamefont {G.}~\bibnamefont
  {Henkelman}}, \bibinfo {author} {\bibfnamefont {A.}~\bibnamefont
  {Arnaldsson}}, \ and\ \bibinfo {author} {\bibfnamefont {H.}~\bibnamefont
  {J\'{o}nsson}},\ }\href@noop {} {\bibfield  {journal} {\bibinfo  {journal}
  {Comput. Mater. Sci.}\ }\textbf {\bibinfo {volume} {36}},\ \bibinfo {pages}
  {254} (\bibinfo {year} {2006})}\BibitemShut {NoStop}%
\bibitem [{\citenamefont {Nartowski}\ \emph {et~al.}(1999)\citenamefont
  {Nartowski}, \citenamefont {Parkin}, \citenamefont {MacKenzie}, \citenamefont
  {Craven},\ and\ \citenamefont {MacLeod}}]{Nartowski-JMC-1999}%
  \BibitemOpen
  \bibfield  {author} {\bibinfo {author} {\bibfnamefont {A.~M.}\ \bibnamefont
  {Nartowski}}, \bibinfo {author} {\bibfnamefont {I.~P.}\ \bibnamefont
  {Parkin}}, \bibinfo {author} {\bibfnamefont {M.}~\bibnamefont {MacKenzie}},
  \bibinfo {author} {\bibfnamefont {A.~J.}\ \bibnamefont {Craven}}, \ and\
  \bibinfo {author} {\bibfnamefont {I.}~\bibnamefont {MacLeod}},\ }\href@noop
  {} {\bibfield  {journal} {\bibinfo  {journal} {J. Mater. Chem.}\ }\textbf
  {\bibinfo {volume} {9}},\ \bibinfo {pages} {1275} (\bibinfo {year}
  {1999})}\BibitemShut {NoStop}%
\bibitem [{\citenamefont {Cowley}(1976)}]{CowleyRA-PRB-1976}%
  \BibitemOpen
  \bibfield  {author} {\bibinfo {author} {\bibfnamefont {R.~A.}\ \bibnamefont
  {Cowley}},\ }\href@noop {} {\bibfield  {journal} {\bibinfo  {journal} {Phys.
  Rev. B}\ }\textbf {\bibinfo {volume} {13}},\ \bibinfo {pages} {4877}
  (\bibinfo {year} {1976})}\BibitemShut {NoStop}%
\bibitem [{\citenamefont {Hill}(1952)}]{HillR-PRSA-1952}%
  \BibitemOpen
  \bibfield  {author} {\bibinfo {author} {\bibfnamefont {R.}~\bibnamefont
  {Hill}},\ }\href@noop {} {\bibfield  {journal} {\bibinfo  {journal} {Proc.
  Phys. Soc. A}\ }\textbf {\bibinfo {volume} {65}},\ \bibinfo {pages} {349}
  (\bibinfo {year} {1952})}\BibitemShut {NoStop}%
\bibitem [{\citenamefont {Pugh}(1954)}]{PughSF-PM-1954}%
  \BibitemOpen
  \bibfield  {author} {\bibinfo {author} {\bibfnamefont {S.~F.}\ \bibnamefont
  {Pugh}},\ }\href@noop {} {\bibfield  {journal} {\bibinfo  {journal} {Philos.
  Mag.}\ }\textbf {\bibinfo {volume} {45}},\ \bibinfo {pages} {823} (\bibinfo
  {year} {1954})}\BibitemShut {NoStop}%
\bibitem [{\citenamefont {Chen}\ \emph {et~al.}(2011)\citenamefont {Chen},
  \citenamefont {Niu}, \citenamefont {Li},\ and\ \citenamefont
  {Li}}]{ChenXQ-IM-2011}%
  \BibitemOpen
  \bibfield  {author} {\bibinfo {author} {\bibfnamefont {X.}~\bibnamefont
  {Chen}}, \bibinfo {author} {\bibfnamefont {H.}~\bibnamefont {Niu}}, \bibinfo
  {author} {\bibfnamefont {D.}~\bibnamefont {Li}}, \ and\ \bibinfo {author}
  {\bibfnamefont {Y.}~\bibnamefont {Li}},\ }\href@noop {} {\bibfield  {journal}
  {\bibinfo  {journal} {Intermetallics}\ }\textbf {\bibinfo {volume} {19}},\
  \bibinfo {pages} {1275} (\bibinfo {year} {2011})}\BibitemShut {NoStop}%
\bibitem [{\citenamefont {He}\ \emph {et~al.}(2008)\citenamefont {He},
  \citenamefont {Lin}, \citenamefont {Wang}, \citenamefont {Bao},\ and\
  \citenamefont {Zhou}}]{HeL-SM-2008}%
  \BibitemOpen
  \bibfield  {author} {\bibinfo {author} {\bibfnamefont {L.}~\bibnamefont
  {He}}, \bibinfo {author} {\bibfnamefont {Z.}~\bibnamefont {Lin}}, \bibinfo
  {author} {\bibfnamefont {J.}~\bibnamefont {Wang}}, \bibinfo {author}
  {\bibfnamefont {Y.}~\bibnamefont {Bao}}, \ and\ \bibinfo {author}
  {\bibfnamefont {Y.}~\bibnamefont {Zhou}},\ }\href@noop {} {\bibfield
  {journal} {\bibinfo  {journal} {Scr. Mater.}\ }\textbf {\bibinfo {volume}
  {58}},\ \bibinfo {pages} {679} (\bibinfo {year} {2008})}\BibitemShut
  {NoStop}%
\bibitem [{\citenamefont {Weber}(1973)}]{WeberW-PRB-1973}%
  \BibitemOpen
  \bibfield  {author} {\bibinfo {author} {\bibfnamefont {W.}~\bibnamefont
  {Weber}},\ }\href@noop {} {\bibfield  {journal} {\bibinfo  {journal} {Phys.
  Rev. B}\ }\textbf {\bibinfo {volume} {8}},\ \bibinfo {pages} {5082} (\bibinfo
  {year} {1973})}\BibitemShut {NoStop}%
\end{thebibliography}%

\end{document}